\documentclass[prd,preprint,showpacs]{revtex4}
\usepackage{graphicx}
\usepackage{amsfonts}
\usepackage{amssymb}
\usepackage{amsmath}
\newcommand{\tmop}[1]{#1}
\newenvironment{itemizedot}
  {\begin{itemize}}{\end{itemize}}
\newcommand{\tmem}[1]{{\em #1\/}}
\newcommand{\tmfloatcontents}{}
\newlength{\tmfloatwidth}
\newcommand{\tmfloat}[5]{
  \renewcommand{\tmfloatcontents}{#4}
  \setlength{\tmfloatwidth}{\widthof{\tmfloatcontents}+1in}
  \ifthenelse{\equal{#2}{small}}
    {\ifthenelse{\lengthtest{\tmfloatwidth > \linewidth}}
      {\setlength{\tmfloatwidth}{\linewidth}}{}}
    {\setlength{\tmfloatwidth}{\linewidth}}  \begin{minipage}[#1]{\tmfloatwidth}
    \begin{center}
      \tmfloatcontents
      \captionof{#3}{#5}
    \end{center}
  \end{minipage}}
\begin{document}
\title{Determining weak phase $\gamma$ and probing new physics in \\ $b
\rightarrow s$ transitions from $B \rightarrow \eta^{(' )} K$}
\author{Yu-Feng Zhou}
\affiliation{Institute for Physics, Dortmund University,D-44221,Dortmund, Germany}

\email[Email: ]{zhou@zylon.physik.uni-dortmund.de}
\date{\today}
\begin{abstract}
  We present a method of determining  weak phase $\gamma$ in the
  Cabibbo-Kobayashi-Maskawa matrix from decays $B \rightarrow
  \eta K, \eta' K$ alone.  
% The method depends on measurements of direct
%  and mixing-induced CP asymmetries in both $\eta' K$ and $\eta K$
%  modes which will be available in the near future.
%
  Given a large ratio between color-suppressed and color-allowed tree
  diagrams  extracted from
  global $\pi\pi(K)$ fits, $\gamma$ is  determined from the current
  data of $\eta' K$ and the result is in agreement with the
  global Standard Model(SM) fits. However, a smaller ratio from
  factorization based calculations gives $\gamma\sim 90^\circ$.
  New physics beyond the SM can be singled out if $\gamma$ obtained in
  $\eta^{(')} K$ modes is significantly different than the ones from
  other modes or other approaches. The effective value of $\gamma$ from $\eta'
  K$ is very sensitive to new physics contributions and can be used to
  extract new physics parameters for a class of models which do not
  give contributions to strong phases significantly.
\end{abstract}
%\preprint{DO-TH/05-XX}
%\preprint{$Revision: 8.1 $}
\preprint{DO-TH 05/08}
%\keywords{}
\maketitle
%%%%%%%%%%%%%%%%%%%%%%%%%%%%%%%%%%%%%%%%%%%%%%%%%%%%%%%%%%%%%%%%%%%%%%%%%%%%%%%%
%%%%%%%%%%%%%%%%%%%%%%%%%%%%-Begin-%%%%%%%%%%%%%%%%%%%%%%%%%%%%%%%%%%%%%%%%%%%%%
%%%%%%%%%%%%%%%%%%%%%%%%%%%%%%%%%%%%%%%%%%%%%%%%%%%%%%%%%%%%%%%%%%%%%%%%%%%%%%%%

%%%%%%%%%%%%%%%%%%%%%%%%%%%%%%%%%%%%%%%%
\section{Introduction}
%%%%%%%%%%%%%%%%%%%%%%%%%%%%%%%%%%%%%%%%
%
Precisely obtaining the weak phase $\alpha$, $\beta$ and $\gamma$ in
the Cabibbo-Kobayashi-Maskawa (CKM) matrix is one of the central
issues in the current studies of $B$ decays.  Besides global fits to
all the indirect measurements in the Standard Model
(SM)\cite{Charles:2004jd,Bona:2005vz} or measurements on the
time-dependent CP asymmetry in $B \rightarrow J / \psi K_S$, the phase
angles $\alpha$ and $\gamma$ of the unitarity triangle can also be
probed from hadronic charmless $B$ decays. In the charmless decay
modes $B \rightarrow P P$ with $P$ denoting a pseudo-scalar final
state, the weak phase $\gamma$ can be determined either with
theoretical inputs such as QCD factorization
\cite{Beneke:2001ev,Beneke:2003zv}, perturbation QCD
\cite{Keum:2000ph,Keum:2002ri,Keum:2002vi} and soft-collinear
effective theories \cite{Bauer:2004tj,Bauer:2004dg,Grossman:2005jb} etc, or through
model independent phenomenological methods based on flavor SU(3)
symmetry
\cite{zeppenfeld:1981ex,Savage:1989jx,Chau:1990ay,Gronau:1994rj,Gronau:1995hn,Gronau:1995ng,Gronau:2002gj,Chiang:2005kz}.

Within the flavor SU(3) symmetry,  direct $B$ decay amplitudes are
described by a set of flavor topological diagrams. The leading
diagrams involve: a tree diagram $\mathcal{T}$, a color suppressed
tree diagram $\mathcal{C}$, a flavor octet (singlet) QCD penguin 
diagram $\mathcal{P} ( \mathcal{S} )$ and a color allowed (color-suppressed)
electroweak penguin diagram $\mathcal{P}_{EW}$ ($\mathcal{P}_{EW}^C$)
etc. The hierarchical structure in the size of these diagrams 
simplifies the analysis and makes it powerful in exploring
the hadronic $B$ decays.  Recent global fits using the
diagrammatic method have already shown that the weak phase $\gamma$
can be determined with a reasonable precision and the obtained value
agrees well with the one from the global CKM fit 
\cite{Wu:2000rb,Zhou:2000hg,He:2000ys,Ali:2004hb,Chiang:2004nm,Wu:2004xx,Wu:2005hi}.

However, the current data also exhibit some puzzling patterns which
needs further understanding. The unexpected large branching ratio of
$\pi^0 \pi^0$ and the relative suppression of $\pi^+ \pi^-$ possess a
big theoretical challenge and may require  large nonfactorizable
contributions\cite{Buras:2003dj,Buras:2004ub,Buras:2004th}; the relative enhancement of $\pi^0 \bar{K}^0$ to $\pi^+
K^-$ may lead to an enhancement of electroweak penguin which could be a
signal of new physics ( see, eg.
\cite{Yoshikawa:2003hb,Mishima:2004um,Buras:2003dj,Buras:2004ub,Buras:2004th%
,Wu:2004xx,Wu:2005hi}).
The recently measured mixing-induced CP asymmetries of ($\omega,
\phi$, $\pi^0, \eta' ) K_S$ though not conclusive yet, suggest a
possibility that the weak phase $\beta$ obtained from $b \rightarrow
s$ penguin-dominant processes may deviate  from the one
determined from $b \rightarrow c$ tree-dominated process $J / \psi K_S$
\cite{Buchalla:2005us,Beneke:2005pu,Kim:2005pe}.

The global fit to all the charmless $B$ decay modes connected by flavor SU(3)
symmetry is the most consistent way to explore the weak phases and the
involved hadronic decay amplitudes. However, to get more insight on the
potential inconsistencies in the theory and a better understanding of the
strong dynamics in hadronic $B$ decays it is  usefully to divide the whole
decay modes into several subsets in which the relevant parameters can be
investigated individually. The comparison among the same quantities obtained
from different subsets will not only provide us important cross-checks but
also shed light on the origins of those puzzles and possible signals of new
physics beyond the SM.

% pi pi modes %
For instance, in $\pi \pi$ system the three decays modes $\pi^+ \pi^-,
\text{$\pi^0 \pi^0 $ and $\pi^0 \pi^-$ }$ provide at most seven
independent observables including three branching ratios, two direct
CP asymmetries (the direct CP asymmetry for $\pi^0 \pi^-$ is predicted
to be vanishing in SM) and two mixing-induced CP asymmetries, enough
to determine the involved hadronic amplitudes $\mathcal{T}$,
$\mathcal{C}$, $\mathcal{P}$ and also the weak phase $\gamma$. In
$\pi \pi$ modes, the electroweak penguins are small and negligible.
The recent fits taking weak phase $\beta$ as input show a good
determination of all the amplitudes.  The ratio of
$\mathcal{C}/\mathcal{T}$ is found to be large close to
0.8\cite{Chiang:2004nm,He:2004ck,Charng:2004ed,Wu:2004xx,Wu:2005hi,Kim:2005jp},
the weak phase $\gamma$ is determined up to a multi-fold ambiguity and
one of them agrees well with the SM global fit value $\sim 62^\circ$.
%
% pi K modes %
In $\pi K$ system, the available data involve four CP averaged
branching ratios, three direct CP asymmetries ( the direct CP asymmetry
in $B^- \rightarrow \pi^- \bar{K}^0$ is predicted to be nearly zero
when annihilation diagram is negligible).  Plussing a mixing-induced CP
asymmetry in $B \rightarrow \pi^0 K_S$, there are eight data points in
total. The independent flavor diagrams include $\mathcal{T}$,
$\mathcal{C}$ and $\mathcal{P}$. The electro-weak penguin
$\mathcal{P}_{\tmop{EW}}$ is significant but can be related to tree
type diagrams in the SM \cite{Neubert:1998re,Gronau:2003kj}. Other
parameters in the CKM matrix elements can be chosen as angles $\gamma$
and $\beta$ or the Wolfenstein parameter $\rho$ and $\eta$.  Thus the
shape of the whole unitarity triangle can be in principle determined
in $\pi K$ modes alone\cite{Imbeault:2003it}. The current data of
$\pi K$ are not enough to perform such an independent determination.
Taking the SM value of weak phase $\gamma$ and $\beta$ as inputs, one
can extract other hadronic amplitudes. The recent fits show a even
larger value of $\mathcal{C}/\mathcal{T} \sim 1.7$ and enhancement of
$\mathcal{P}_{\tmop{EW}}$\cite{Baek:2004rp,Wu:2005hi}.

% eta K modes %
In the present paper, we discuss the determination of $\gamma$ from an
other important subset, the $\eta^{(')} K$ modes. The advantages of
using $\eta^{(' )} K$ final states over the $\pi\pi$ and $\pi K$
states are as follows
\begin{itemizedot}
\item All the four $\eta^{(' )} K$ modes are penguin dominant with
  appreciable tree-penguin interferences. Nonvanishing direct CP
  asymmetries are expected in {\tmem{all}} the four decay modes, while
  in $\pi \pi$($\pi K$) one of the direct CP asymmetry in
  $\pi^-\pi^0$($\pi^-\bar{K}^0$) is predicted to be nearly zero.
  
\item The two neutral modes in $\eta^{(')}K$ will provides two
  additional data points from mixing-induced CP asymmetries in
  $\eta^{(' )} K_S$, while in $\pi K$ there is only one.
  
\item Most importantly, the flavor topological structure in
  $\eta^{(')}K$ amplitudes allows a regrouping of penguin type
  diagrams in such a way that the number of independent hadronic
  amplitudes can be reduced to four complex parameters.
  
\item The electroweak penguin diagram $\mathcal{P}_{EW}$ can be included
  in the reduced hadronic parameters. It is not necessary to assume
  the SM relation between electroweak penguin and tree type diagram.
  This is of particular importance as the current data imply the
  possibility of new physics beyond the SM.

\end{itemizedot}

Thus in $\eta^{(' )} K$ modes there will be at most ten observables
available, enough to simultaneously determine all the involved
diagrammatic amplitudes, the weak phase $\gamma$ and $\beta$ which
determine the apex of the unitarity triangle. This method
distinguishes itself from the previous ones in that it makes use of the
$\eta^{(')} K$ modes {\tmem{alone}} while the previous methods focus on
constructing quadrangles connecting to $\pi K$ modes using SU(3)
symmetry\cite{Gronau:1995ng,Dighe:1995bm}. 

This paper is organized as follows, in section II, we present details
of determining weak phase $\gamma$ from $\eta^{(' )} K$ modes.  In
section III, the implications from the current data of $\eta' K$ is
discussed. We take typical values of hadronic parameters as inputs to
constrain $\gamma$ from $\eta' K$ modes.  In section IV, the new
physics effects on the $\gamma$ determination is discussed. We finally
conclude in section V.

%
%%%%%%%%%%%%%%%%%%%%%%%%%%%%%%%%%%%%%%%%%%%%%%%%%%%%%%%%%
\section{Determining $\gamma$ from $\eta^{(')}K$}
%%%%%%%%%%%%%%%%%%%%%%%%%%%%%%%%%%%%%%%%%%%%%%%%%%%%%%%%%
%
We assume flavor SU(3) symmetry and take the following diagrammatic
decomposition for $B\to \eta^{(')}K$ decay amplitudes
\cite{Dighe:1995gq}.
\begin{eqnarray}
  \bar{\mathcal{A}} ( \eta \bar{K}^0 ) & = &  \frac{1}{\sqrt{3}} \left(
  \mathcal{C}+\mathcal{P}_{\eta} \right) , \nonumber\\
  \bar{\mathcal{A}} ( \eta' \bar{K}^0 ) & =& \frac{1}{\sqrt{6}} \left(
  \mathcal{C}+\mathcal{P}_{\eta'} \right) , \nonumber\\
  \bar{\mathcal{A}} ( \eta K^- ) & = &  \frac{1}{\sqrt{3}} \left(
  \mathcal{T}+\mathcal{C}+\mathcal{P}_{\eta} \right) , \nonumber\\
  \bar{\mathcal{A}} ( \eta' K^- ) &= & \frac{1}{\sqrt{6}} \left(
  \mathcal{T}+\mathcal{C}+\mathcal{P}_{\eta'} \right) ,
\end{eqnarray}
which corresponds to the flavor contents of $\eta = ( -s
\bar{s} + u \bar{u} + d \bar{d} \text{} ) / \sqrt{3}$ and $\eta' = ( 2
s \bar{s} + u \bar{u} + d \bar{d} \text{} ) / \sqrt{6}$ respectively.
This is in accordance with an $\eta_8 - \eta_0$ mixing angle of $\theta =
\arcsin ( -1 / 3 ) \simeq - 19.5^{\circ}$\cite{Dighe:1995gq}. Such a simple
mixing scheme is a good approximation in phenomenology and is extensively used
in the recent analyses of hadronic $B$ and $D$ decays 
\cite{Dighe:1997hm,Fu:2003fy,Chiang:2002mr,Wu:2004ht}. 
The two penguin type diagram are given by
\begin{align}%\label{}
\mathcal{P}_{\eta} &\equiv\mathcal{S}+ \frac{2}{3} \mathcal{P}_{\tmop{EW}},
\nonumber\\
\mathcal{P}_{\eta'} &\equiv\ 3\mathcal{P}+ 4\mathcal{S}- \frac{1}{3}
\mathcal{P}_{\tmop{EW}}.
\end{align}
In the above expressions we assume that the color-suppressed
electro-weak penguin $\mathcal{P}_{\tmop{EW}}^C$ and annihilation
diagram $\mathcal{A}$ are small and negligible. We shall also assume
the $t$-quark dominance in the penguin diagrams.  With these
assumptions, all the decay amplitudes depend on four complex parameters
$\mathcal{C}, \mathcal{T}, \mathcal{P}_{\eta} \mbox{ and }
\mathcal{P}_{\eta'}$.
%
% mixing dependence
%
%
%Note that such a recombination is independent of the details of $\eta -
%\eta'$ mixing.  In a more general mixing scheme with two-mixing angles
%\cite{Leutwyler:1998yr,Feldmann:1999uf} one still gets four independent amplitudes with the definitions of
%penguin type diagrams change to
%
%\begin{align}%\label{}
%\mathcal{P}_{\eta}&=(\cos\theta_8/\sqrt{2}+2\sin\theta_0) \mathcal{P}
%-3\sin\theta_0\mathcal{S}+\cos\theta_8 \mathcal{P}_{EW}/\sqrt{2} ,
%\nonumber\\
%\mathcal{P}_{\eta'}&=-(\sin\theta_8/-2\sqrt{2}\cos\theta_0) \mathcal{P}
%+3\sqrt{2}\cos\theta_0\mathcal{S}+\sin\theta_8 \mathcal{P}_{EW} ,
%\end{align}
%
%where $\theta_0\simeq -9.2^\circ$ and $\theta_8\simeq -21.2^\circ$ are
%the octet and singlet mixing angles. The tree-type diagrams are simply
%multiplied by global mixing angle dependent factors. 
%
%
The two weak
phases $\gamma$ and $\beta$ enter the expressions from direct and
mixing-induced CP asymmetries as additional free parameters.  Removing
a overall strong phase, there are 9 real free parameters to be
determined by 10 observables in $B \rightarrow \eta^{(' )} K$ modes
which include four CP averaged decay rates, four direct CP asymmetries
and two mixing-induced CP asymmetries in $\eta^{(' )} K_S$.
%
% new mixing dependence
%
Although the expressions of $P_{\eta}$ and $P_{\eta'}$ depends
on $\eta-\eta'$ mixing scheme, the isospin symmetry guarantees
that neutral($\eta(\eta')\bar{K}^0$) and charged ($\eta(\eta')K^-$) modes
have the same coefficients for $\mathcal{S},\mathcal{P}$ and also 
$\mathcal{P}_{EW}$, which allows the reduction to a single
penguin type parameter. Thus the number of free parameters is the same
 for  other mixing schemes
such as FKS and two-mixing angle schemes ( see, e.g. 
\cite{Leutwyler:1998yr,Feldmann:1999uf}).
%For a detailed discussion of $\eta-\eta'$ mixing effects in 
%charmless $B$ decays, we refer to Ref.\cite{Beneke:2002jn}.

The CP averaged branching ratio is
defined through
\begin{eqnarray}
  \tmop{Br} & \equiv & \frac{1}{2} \tau ( | \bar{\mathcal{A}} |^2 + |\mathcal{A}|^2
  ) ,
\end{eqnarray}
where the factor $\tau$ stands for the life time difference in $B$ mesons and is
normalized to  $\tau = 1 ( \tau_+ / \tau_0 )$ for neutral(charged) modes with
$\tau_0(\tau_+)$ the life time for neutral (charged) $B$ mesons
and $\tau_+/\tau_0=1.086$.
The definition of direct CP asymmetry is
\begin{eqnarray}
  a_{\tmop{cp}} & \equiv & \frac{| \bar{\mathcal{A}} |^2 - |\mathcal{A}|^2}{|
  \bar{\mathcal{A}} |^2 + |\mathcal{A}|^2} .
\end{eqnarray}
The mixing-induced CP asymmetry is defined as
\begin{eqnarray}
  a_{\tmop{cp}} ( t ) & \equiv & \frac{\tmop{Br} ( \bar{B}^0 ( t ) \rightarrow f )
  - \tmop{Br} ( B^0 ( t ) \rightarrow f )}{\tmop{Br} ( \bar{B}^0 ( t )
  \rightarrow f ) + \tmop{Br} ( B^0 ( t ) \rightarrow f )}
\nonumber\\
  & = &S \sin(\Delta m_B t)- C\cos (\Delta m_B t) ,
\end{eqnarray}
where
\begin{eqnarray}
  S = \frac{\tmop{Im} \lambda}{| \lambda |^2 + 1} & , & \mbox{ and } \lambda = -
  e^{- 2 i \phi_d}  \frac{\bar{\mathcal{A}}}{\mathcal{A}} ,
\end{eqnarray}
with $\phi_d$  the weak phase appearing in $B^0 - \bar{B}^0$ mixing and $\phi_d =
\beta$ in the SM. The coefficient $C$ is related to the direct CP asymmetry by
$C=-a_{cp}$. The latest data involving $B \rightarrow \eta^{(' )} K$ are
summarized in Tab.\ref{data} \cite{Aubert:2005iy,Abe:2004xp,hfag}

\begin{table}[htb]\begin{center}
\begin{ruledtabular}
\begin{tabular}{ccccc}
  & CLEO & BaBar & Belle & WA\\
  \hline
  $\tmop{Br} ( \eta \bar{K}^0 )$ & $< 9.3$ & $< 2.5$ & $< 2.0$ & $< 2.0$\\
  \hline
  $\tmop{Br} ( \eta K^- )$ & $2.2^{+ 2.8}_{- 2.2}$ & $3.3 \pm 0.6 \pm 0.3$ &
  $2.1 \pm 0.6 \pm 0.2$ & $2.6 \pm 0.5$\\
  \hline
  $\tmop{Br} ( \eta'  \bar{K}^0 )$ & $89^{+ 18}_{- 16} \pm 9$ & $67.4 \pm 3.3
  \pm 3.3$ & $68 \pm 10^{+ 9}_{- 8}$ & $68.6 \pm 4.2$\\
  \hline
  $\tmop{Br} ( \eta' K^- )$ & $80^{+ 10}_{- 9} \pm 7$ & $68.9 \pm 2 \pm 3.2$ &
  $78 \pm 6 \pm 9$ & $70.8 \pm 3.4$\\
  \hline
  $a_{\tmop{cp}} ( \eta K^- )$ &  & $- 0.2 \pm 0.15 \pm 0.01$ & $- 0.49 \pm
  0.31 \pm 0.07$ & $- 0.25 \pm 0.14$\\
  \hline
  $a_{\tmop{cp}} ( \eta'  \bar{K}^0 )$ &  &  &  & $(0.04 \pm 0.08)$\\
  \hline
  $a_{\tmop{cp}} ( \eta' K^-$) & $0.03 \pm 0.12 \pm 0.02$ & $0.033 \pm 0.028
  \pm 0.005$ & $- 0.015 \pm 0.007 \pm 0.009$ & $0.027 \pm 0.025$\\
  \hline
  $S'$ &  & $0.30 \pm 0.14 \pm 0.02$ & $0.65 \pm 0.18 \pm 0.04$ & $0.43 \pm
  0.11$\\
\end{tabular}
\end{ruledtabular}
\end{center}
\caption{The latest world average of branching ratios ( 
in unit of $10^{-6}$), direct CP violations as well as mixing-induced CP violations for $B \rightarrow
\eta K, \eta' K$ modes. The direct CP asymmetry of $\eta' \bar{K}^0$ comes
from time-dependent CP asymmetry measurements of $\eta' K_S$.
}
\label{data}
\end{table}

%\begin{table}[htb]\begin{center}
%\begin{ruledtabular}
%\begin{tabular}{ccccc}
%  & CLEO & BaBar & Belle & WA\\
%  \hline
%  $\tmop{Br} ( \eta \bar{K}^0 )$ &  &  &  & $1.5 \pm 0.6$\\
%  \hline
%  $\tmop{Br} ( \eta K^- )$ &  &  &  & $2.6 \pm 0.5$\\
%  \hline
%  $\tmop{Br} ( \eta'  \bar{K}^0 )$ &  &  &  & $65.2^{+ 6.0}_{- 5.9}$\\
%  \hline
%  $\tmop{Br} ( \eta' K^- )$ &  &  &  & $77.6^{+ 4.6}_{- 4.5}$\\
%  \hline
%  $a_{\tmop{cp}} ( \eta K^- )$ &  &  &  & $- 0.25 \pm 0.14$\\
%  \hline
%  $a_{\tmop{cp}} ( \eta'  \bar{K}^0 )$ &  &  &  & $0.04 \pm 0.008$\\
%  \hline
%  $a_{\tmop{cp}} ( \eta' K^-$) &  &  &  & $0.027 \pm 0.025$\\
%  \hline
%  $S'$ &  &  &  & $0.41 \pm 0.11$\\
%\end{tabular}
%\end{ruledtabular}
%\end{center}
%\caption{The latest world average of branching ratios,
%direct CP violations as well as mixng-incuded CP violations for $B \rightarrow
%\eta K, \eta' K$ modes.}\label{data}
%\end{table}

It is well known that the unusually large branching ratios of $B
\rightarrow \eta' K$ modes may require a enhancement of flavor singlet penguin
diagrams $\mathcal{S}$, which possess an other theoretical challenge and
is still under extensive theoretical study 
( see, e.g.\cite{etaK}).  
%
% QCD factorization 
%
The flavour singlet contribution can be systematically calculated
in QCD factorization, the results favour a smaller value with
significant theoretical uncertainties\cite{Beneke:2002jn}. 
However, for the
purpose of extracting weak phases one needs only the ratios of
decay rates between 
neutral and charged modes in which the penguin amplitudes cancel
in a great extent, making the results in sensitive to $\mathcal{S}$. 
We then define a ratio between neutral and charged decay rates as
\begin{eqnarray}
  R^{(' )} & \equiv & \frac{\tau_+}{\tau_0}  \frac{\tmop{Br} ( \eta^{(' )}
  \bar{K}^0 )}{\tmop{Br} ( \eta^{(' )} K^- )} ,
\end{eqnarray}
The current data of $B \rightarrow \eta' K$ gives
\begin{eqnarray}
  R' = 1.04 \pm 0.08.
\end{eqnarray}
%and
%\begin{eqnarray}
%  R <0.77 .
%\end{eqnarray}
%
%It will be seen below that the uncertainty in $R'$ is the dominant 
%source of uncertainty in the estimation of $\tan\gamma$ as it is very
%close to unity.  
The corresponding ratio in $\eta K$ modes gives $R<0.83$. 
The ratio between tree and penguin type diagrams are
parameterized as
\begin{eqnarray}
  \zeta^{(' )} e^{i \delta^{(' )}} = \frac{\mathcal{C}}{\mathcal{P}_{\eta^{('
  )}}}e^{i\gamma} & , & \chi^{(' )} e^{i \omega^{(' )}} =
  \frac{\mathcal{T}+\mathcal{C}}{\mathcal{P}_{\eta^{(' )}}}e^{i\gamma} , 
\end{eqnarray}
where $\zeta^{(' )}$ and $\chi^{(' )}$ are both real-valued.
$\delta^{(')}$ and $\omega^{(')}$ are purely strong phases as the weak
phase $\gamma$ has been extracted from the definitions. We further
define a ratio between color-suppressed and color-allowed tree
diagrams
\begin{eqnarray}
  r e^{i \varphi} \equiv \frac{\zeta^{(' )} e^{i \delta^{(' )}}}{\chi^{(' )}
  e^{i \omega^{(' )}}} & = & \frac{\mathcal{C}}{\mathcal{T}+\mathcal{C}} ,
\end{eqnarray}
with $r = | \zeta^{(' )} / \chi^{(' )} |$ and $\varphi = \delta^{(' )}
- \omega^{(' )}$, which are common to both $\eta K$ and $\eta' K$ modes.
%If one uses only the ratios among decay rates, there are eight free
%parameters to be determined by 8 observables.

All the parameters can be solved numerically from the above equations.
They can also be solved analytically to the leading order expansion of
$\zeta^{(' )}$ and $\chi^{(' )}$. Taking the $\eta' K$ modes as an
example, to the leading order of $\zeta'$ and $\omega'$, the ratio of
the decay rates is given by
\begin{eqnarray}
  R' \equiv 1+\Delta R' & \simeq &1+ 2 \zeta' \left[ \cos \delta' - \frac{1}{r} \cos
  ( \delta' - \varphi ) \right] \cos \gamma .  \label{R}
\end{eqnarray}
The two direct CP asymmetries are
\begin{align}
  a'_0 \equiv a_{\tmop{cp}} ( \eta' \bar{K}^0 ) & \simeq  2 \zeta' \sin \delta'
  \sin \gamma, \\
  a_-' \equiv a_{\tmop{cp}} ( \eta' K^- ) & \simeq  2 \frac{\zeta'}{r} \sin (
  \delta' - \varphi ) \sin \gamma .  \label{acp}
\end{align}
The mixing-induced CP violation is found to be
\begin{eqnarray}
  S'\equiv S(\eta'K_S) & \simeq &  \text{$\sin 2 \beta$+} 2 \zeta' \cos 2 \beta \cos \delta' \sin
  \gamma .  \label{S}
\end{eqnarray}
In the above expressions, we use the primed quantities such as $R'$,
$a'_0$, $a'_-$ and $S'$ to denote ratio of decay rates, direct
and mixing-induced CP asymmetries respectively in $\eta' K$ modes. For
$B \rightarrow \eta K$ process, equations similar to
Eq.(\ref{R})-(\ref{S}) can be constructed with the substitution of
primed quantities to be unprimed ones i.e $R$, $a_0$, $a_-$ and $S$
etc.  The Eqs.(\ref{R})-(\ref{S}) together with the ones for $\eta K$
modes provide eight equations which constrain the eight parameters,
$$\zeta', \ \delta', \ \zeta, \ \delta, \ r,\ \varphi, \ \gamma \
\mbox{and} \ \beta .$$ 
A simultaneous determination of
$\gamma$ and $\beta$ will allow a reconstruction of the unitarity
triangle from $\eta^{(' )} K$ modes alone.

Since great success has already been achieved in the measurement of
$\sin 2 \beta$ from $B \rightarrow J / \psi K_S$ in the two
$B$-factories and the value obtained agrees remarkably with the one
from global fits to all the indirect measurements such as neutral $B$
and $K$ meson mixing and semileptonic $B$ decays etc, throughout this
paper we shall take the value of \cite{Aubert:2002ic,Abe:2002bx}
\begin{align}%\label{}
\sin 2 \beta = 0.687 \pm 0.032
\end{align}
from $B \rightarrow J / \psi K_S$ as input and focus on the
determination of the less known weak phase $\gamma$ in $\eta^{(' )} K$
modes.

Following this strategy, the value of $r$ and $\varphi$ are
determined purely by direct CP asymmetries and $\beta$
\begin{eqnarray}
  \tan \varphi & \simeq & 
\frac{(a'_0 a_- - a_0 a'_-) \cos2\beta}
{a_-(S'-\sin2\beta)- a'_-( S-\sin2\beta)} , 
\end{eqnarray}
and
\begin{eqnarray}
  r & \simeq & 
\frac{a'_0}{a'_-}\left( \cos\varphi - \frac{S'-\sin2\beta}{a'_0 \cos2\beta}\sin\varphi
\right) .
\end{eqnarray}
Note that the determination of $\varphi$ requires CP asymmetry
measurements for both $\eta K$ and $\eta' K$ modes. 
The solution to  $\gamma$ in terms of  of $r$, $\varphi$ and $\beta$ is
found straight forwardly
\begin{eqnarray}
  \tan \gamma & \simeq & \frac{1}{r \Delta R'} \left[ ( r - \cos \varphi ) \frac{S'
  - \text{$\sin 2 \beta$}}{\cos 2 \beta} - a'_0 \cdot \sin \varphi \right] .
  \label{gamma}
\end{eqnarray}
Thus $\gamma$ is  determined up to discrete ambiguities.  The above
expression forms the base of the present paper.  The weak phase
$\gamma$ does not depend on the ratio $\zeta'$ and $\chi'$.  It only
depends on the ratio between tree type diagrams $r$ and $\varphi$.  The accuracy of
$\gamma$ depends heavily on the CP violation measurements. It also
depends on the ratios of the decay rates $R'$. Note again that in this
method the weak phase $\gamma$ is determined within a closed subset of
$\eta^{(' )} K$. No measurements from other modes are needed.
In a typical  case where $\varphi$ is small  and $r< \cos\varphi$,
the second term in the right handed side of Eq.(\ref{gamma}) is
negligible, the sign of $\tan\gamma$ depends on the sign of
$(S'-\sin2\beta)/\Delta R'$.  Thus a positive $\tan\gamma$ nontrivially requires
$R'>1$ since the current data prefer $S'-\sin2\beta<0$.
The value of $\tan\gamma$ will be enhanced if  $r$ or $\Delta R'$
is very small.

%which provides a unique opportunity to investigate the
%interference between weak and strong interactions in these modes. Some
%interesting observations can be found in the following extreme cases:
%%
%\begin{itemize}
%\item[a)] If the future measurements confirm that $S'$ is at it's
%  ``expected'' value of $S'\approx\sin 2\beta$, one finds $\tan \gamma
%  \simeq - a'_0 \cdot \sin \varphi / ( r \Delta R'$). Thus if the
%  current value of $\gamma \simeq 62^{\circ}$ is correct, the strong
%  phase of $\varphi$ must be significantly nonzero as it is enhanced
%  by factor of $1/(r \Delta R')$ which could reach
%  $\mathcal{O}(10^2)$ for a small $r \alt 0.3$.

%\item[b)] If $\varphi \approx 0$ as indicated from the naive
%  factorization calculations, one gets a relation $S' - \sin 2 \beta =
%  r \Delta R' \cos 2 \beta \tan \gamma / ( r - 1 )$. Thus the
%  deviation of $S'$ from $\sin 2 \beta$ is suppressed by $r\Delta R'$.
%  This is in agreement with native factorization estimates. 

%\item[c)] For a small $\varphi$, the second term in Eq.(\ref{gamma})
%  is negligible.  The sign of $\tan\gamma$ depends on
%  $(r-\cos\varphi)(S'-\sin2\beta)$. To get a positive $\tan\gamma$, r
%  must be less than $\cos\varphi$.
%\end{itemize}
%
%  SU(3) breaking
%
The main source of the uncertainties comes from the SU(3) breaking
between $\eta K$ and $\eta' K$ decay amplitudes. At present there is
no robust estimates for SU(3) breaking effects. In the naive
factorization approach the SU(3) breaking arises from two difference
pieces in $B\to \eta^{(')} K$ amplitudes.  One is proportional to the
form factors
$F^{B\to\eta}_0(m^2_B-m^2_\eta)/F^{B\to\eta'}_0(m^2_B-m^2_{\eta'})\approx1.16$
with $F^{B\to \eta(\eta')}$ the form factor of $B\to \eta^{(')}$
transition.  The other one is proportional to the decay constants
$f^u_\eta/f^u_{\eta'}\approx 1.22$.  This is gives an estimate that
the SU(3) breaking effect is up to $\sim 20\%$.  
It needs to be emphasised that $P_{\eta}$ and $P_{\eta'}$ are treated
as two independent parameters not related by SU(3) symmetry. Note that
the SU(3) symmetry could be broken in a more complicated way in the
strong phase\cite{Wu:2002nz} and radiative corrections may give
contributions not proportional to the decay constants
\cite{Beneke:2002jn}.
%
% multi-fold ambiguity
%
  The accuracy of $r$, $\varphi$ and $\delta^{(' )}$ lies on the
  precision of $a_{\tmop{cp}}$s to be measured from $\eta^{(' )} K$
  modes. The branching ratios for $\eta' K$ are known to be large ( a
  few $\times 10^{- 5}$), while the  $\eta K$ modes
  are expected to be an order of magnitude smaller due to it's flavor
  structure \cite{Lipkin:1991us}. However, in the $\eta K$ modes the tree-penguin
  interferences could be stronger and  the direct CP
  asymmetries could be more significant. With the increasing
  statistics in the two $B$-factories, the precision of $a_{\tmop{cp}}
  ( \eta' K$) will be improved.  Higher precision measurements can be
  achieved in the future super-$B$ factories \cite{Hewett:2004tv}.

%%%%%%%%%%%%%%%%%%%%%%%%%%%%%%%%%%%%%%%%%%%%%%%%%%%%
\section{Implications from the latest data}
%%%%%%%%%%%%%%%%%%%%%%%%%%%%%%%%%%%%%%%%%%%%%%%%%%%%

The weak phase $\gamma$ obtained from $\eta^{(' )} K$ modes can be
compared with the one from other methods. The difference, if exists
will shed light on the nonstandard contributions or possible new
physics. At present, the data of the direct and mixing-induced CP
asymmetries for $\eta \bar{K}^0$ are not yet available, one can not
have a practical estimate of $\gamma $ from Eq.(\ref{gamma}).
However, $r$ and $\varphi$ can be extracted from other modes or
calculated theoretically.  Taking $r$ and $\varphi$ as inputs, one can
infer the value of $\gamma$ from $\eta' K$ modes using the current
data and compare it with the SM fit value.  For illustrations, we
consider two typical sets for the value of $r$ and $\varphi$
\begin{itemize}
\item[a)] The values of $r$ and $\varphi$ are extracted from global $\pi \pi$
  and $\pi K$ fit based on flavor SU(3) symmetry. All the recent fits prefer a large
$\mathcal{C}$ 
\cite{Chiang:2003pm,Wu:2004xx,Wu:2005hi,He:2004ck,Charng:2004ed,Kim:2005jp}. 
From an up to date fit in Ref.\cite{Wu:2005hi}, one finds the following values 
  \begin{eqnarray}
  r = 0.56 \pm 0.05 & , & \varphi = (-33.2\pm 6.3)^\circ .  \label{big-r}
  \end{eqnarray}
  The large $r$ is driven by the observed large branching ratio of
  $\pi^0\pi^0$.  The value of $r$ obtained in the $\pi\pi$ and $\pi K$
  fits can be directly used in $\eta' K$ as the leading SU(3) breaking
  effects cancel in the  ratio between $\mathcal{C}$ and $\mathcal{T}$.

\item[b)] The values of $r$ and $\varphi$ are taken from QCD
  factorization calculations \cite{Beneke:2003zv,Beneke:2005vv}, which
  prefers smaller values with considerable uncertainties. In numerical
  estimations we take the following typical values from the latest QCD
  factorization estimate \cite{Beneke:2005vv} 

	\begin{eqnarray} r =
  	0.20\pm0.14 & , & \varphi =-(12\pm18)^\circ .  \label{small-r}
  	\end{eqnarray}
\end{itemize}

%%%%%%%%%%%%%%%%%%%%%%%%%% Fig.1 %%%%%%%%%%%%%%%%%%%%%%%%%%%%%%%%%%%%%%%%%
\begin{figure}[htb]
\begin{center}
  \includegraphics[width=0.7\textwidth]{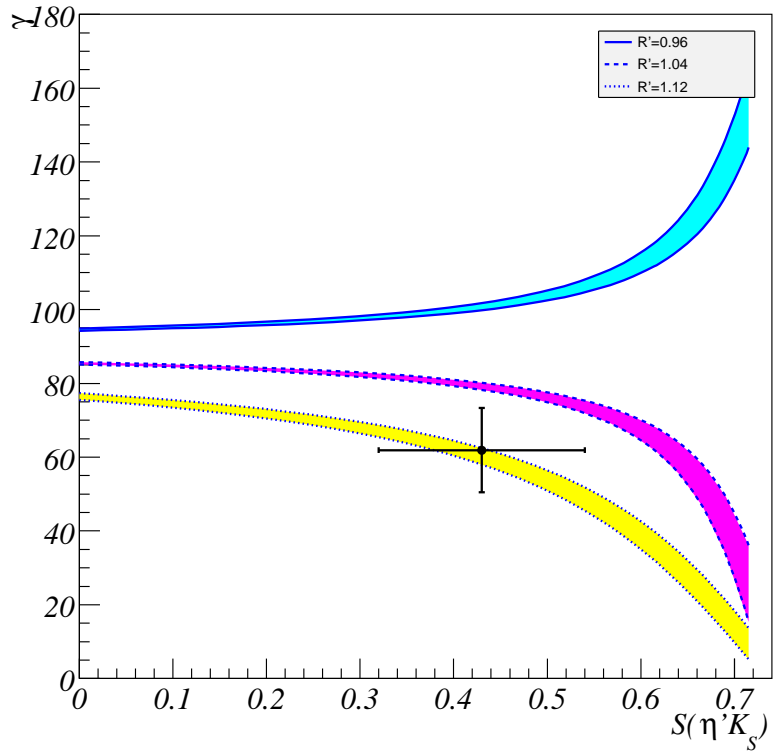}
\end{center}
\caption{Weak phase $\gamma$(in degree) as function of $S'$. The solid,
  dashed and dotted bands corresponds to $R'=0.96,1.04$, and $1.12$
  respectively.  The cross indicates the current experimental
  measurements with the horizontal bar representing the  data of
  $S'$ and the vertical one representing the favored range of $\gamma$
  from global SM fit. The values of $r$ and $\varphi$ are taken from
  Eq.(\ref{big-r}) with uncertainties taken into
  account. }
\label{Fig-1}
\end{figure}
%%%%%%%%%%%%%%%%%%%%%%%%%%%%%%%%%%%%%%%%%%%%%%%%%%%%%%%%%%%%%%%%%%%%%%%%5

In Fig.\ref{Fig-1}. we plot $\gamma$ as a function of $S'$, taking
Eq.(\ref{big-r}) as inputs for three different values of $R'$=0.96,
1.04 and 1.12 respectively, corresponding to the $1\sigma$ allowed
range. 
The figure shows a strong dependence of $\gamma$ on both $S'$ and
$R'$.  For $R'<1$, $\gamma$ grows up with $S'$ increasing
and is always larger than the best fitted value from global CKM fit.
For $R'>1$ , it moves down to the opposite direction and reaches $\sim
60^\circ$ for $S'\simeq 0.5$. For $R'=1$, $\tan\gamma$ becomes
infinity which fix $\gamma$ at $\sim 90^\circ$. The current data of $R'$
can not definitely tell us if $R'$ is greater or smaller than unity.
To have a robust conclusion, higher precision data are urgently
needed.

From Fig.\ref{Fig-1}, one finds a overall consistency with the global
SM fit. For $S'$ and $R'$ varying in the $1\sigma$ range,
the value of $\gamma$ is found to be
\begin{equation}
  45^{\circ} \lesssim \gamma \lesssim 110^{\circ} .
\end{equation}
The error is still significant and the center value gives a slightly 
large $\gamma \sim 78^\circ$.
% comments 
Note that some previous analyses found problems to coincide with a
small $S'$\cite{Chiang:2004nm,Gronau:2005gz}. The difference mainly originates from the data
used in the fits. In the present paper, we use the updated data while
in the previous ones the old data of $Br(\eta' K^-)=77.6\pm 4.6$ and
$Br(\eta' \bar{K}^0)=65.2\pm6.2$ are used which corresponds to
$R'\simeq 0.91$. As it is shown in Fig.\ref{Fig-1}, a small $R'<1$
will not make a good fit.

\begin{figure}[htb]
\begin{center}
\includegraphics[width=0.7\textwidth]{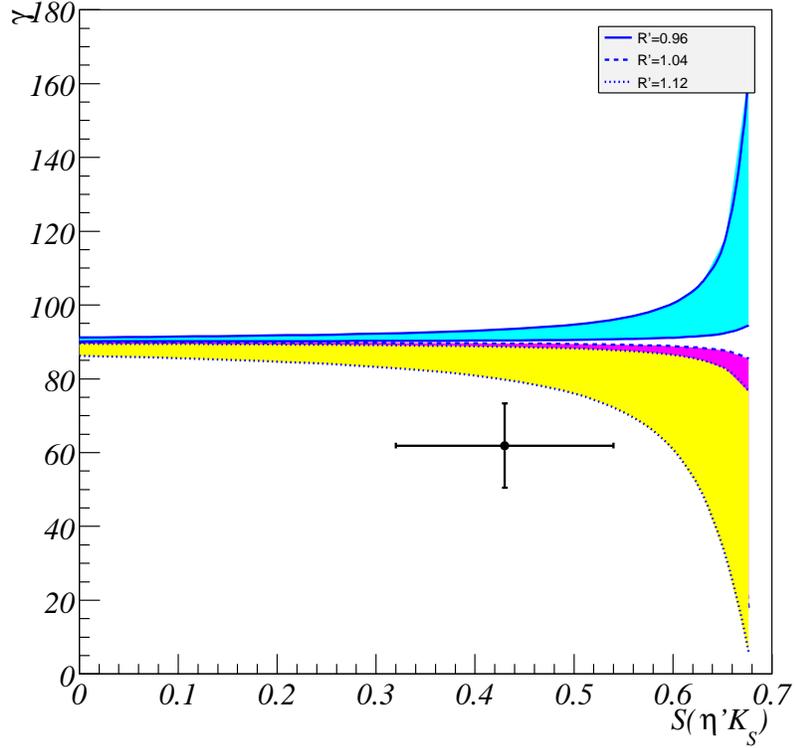}
\end{center}
\caption{ The same as Fig.\ref{Fig-1} while the value of 
$r$ and $\varphi$ are taken from Eq.(\ref{small-r})}\label{Fig-2}
\end{figure}
%%%%%%%%%%%%%%%%%%%%%%%%%%%%%%%%%%%%%%%%%%%%%%%%%%%%%%%%%%%%%%%%%

In Fig.\ref{Fig-2}, a similar plot is made with the values of $r$ and $\varphi$
taken from Eq.(\ref{small-r}). Comparing with Fig.\ref{Fig-1}, one sees 
a smoother dependence  on $R'$ and $S'$, for a smaller $r=0.20$
from Eq.({\tmem{{\tmem{\ref{small-r}}}}}) and 
$R'$, $S'$ in the $1\sigma$ range, the value of $\gamma$ is
found to be confined in a narrow range of
\begin{eqnarray}
  85^{\circ} \lesssim \gamma \lesssim 95^{\circ} &  & 
\end{eqnarray}
Clearly, a large $\gamma\sim 90^\circ$ is favored in this case. The
reason is that the smaller $r$ enhances $\tan\gamma$, making the three
curves closing to each other and forcing $\gamma$ to be $\sim
90^\circ$. In this case $\gamma$ can reach $\sim 60^\circ$ only for
$S'= 0.6\sim 0.7$, i.e. close to the $\sin2\beta$ from $B\to J/\psi
K_S$.  One has to bear in mind that the measurement on $S'$ is not
very conclusive yet as there still exist discrepancy between Babar and
Belle results \cite{Aubert:2005iy,Abe:2004xp}.  Using the PDG average
method, the error should be enlarged by a factor of $\sqrt{\chi^2}$
which is the square root of the chi-square value of the average. This
gives $S'=0.43\pm 0.17$.  However, a large $\gamma$ is still favored
in the enlarged region.  The theoretical prediction to $S'$ based on
QCD factorization prefer that $S'$ is slightly greater than
$\sin2\beta$, $S'-\sin2\beta \simeq
0.01$\cite{Buchalla:2005us,Beneke:2005pu}. This remains to be tested
in the future experiment.

It follows from the above results that if $\gamma$ is indeed around
$62^\circ$, a large $r$ is favored by the current data of $\eta' K$
only, which is independent of the data of $\pi\pi$ and $\pi K$.
Independent determination of the relative size of the color-suppressed
tree diagram may provide us important hints on it's origin. A possible
explanation is that the extracted $\mathcal{C}$ is an effective
amplitude involving other important contributions such as: a large
nonfactorizable $W$-exchanging diagram
$\mathcal{E}$\cite{Buras:2003dj,Buras:2004ub,Buras:2004th}, a large
penguin type diagram contribution through internal $c\bar{c}$ loops,
i.e. the charming penguin \cite{Ciuchini:1997hb,Ciuchini:2001gv},
large final state interactions \cite{Barshay:2004hb,Cheng:2004ru} etc.
The exchange diagram $\mathcal{E}$ only contributes to $\pi\pi$ modes
and will not affect $\pi K$ and $\eta' K$. The charming penguin always
come together with the ordinary penguin diagrams. But the tree-penguin
interferences are different in $\pi\pi,\pi K$ and $\eta' K$.  One can
not expect a universal enhancement pattern of $\mathcal{C}$ in all
modes.  The final state interaction is more process-dependent. Thus if
the ratio $r$ can be precisely determined independently from various
subsets, it is possible to distinguish some of the explanations. For
instance, if large $r$ is confirmed in all the $\pi\pi$, $\pi K$ and
$\eta^{(')}K$ modes, the first explanation will not be favored.

%%%%%%%%%%%%%%%%%%%%%%%%%%%%%%%%%5
\section{New physics effects}
%%%%%%%%%%%%%%%%%%%%%%%%%%%%%%%%%%
%
We proceed to discuss the new physics contributions.  When the weak
phase $\beta$ is taken as known from $B\to J/\psi K_S$, there are
eight data points to constrain seven real parameters in $\eta^{(' )}
K$ system.  The nonzero degree-of-freedom allows one to make
cross-checks for consistency or explore new physics contributions.

The new physics may affect the observables in two different ways. One
is through modifying $B^0-\bar{B}^0$ mixing which makes $\phi_d \neq
\beta$. The consequence is that the mixing induced CP asymmetry for
all the modes will be affected in the same manner, which is not very
likely as the measurement of $\sin2\beta$ from $J/\psi K_S$ agrees
remarkably with all the indirect measurements of the unitarity
triangle and so far no systematic deviations of $\sin2\beta$ from it's
global SM fit value are confirmed in other modes. The other way is that
new physics contributes to decay amplitudes, most likely through $b\to
s$ loop processes. In this case the modifications to direct and
mixing-induced CP asymmetries will be process dependent.

Taking the $\eta'K$ modes as an example, 
we parameterize the new physics contribution to $b \rightarrow s$ penguin in the
following form
\begin{eqnarray}
  \bar{\mathcal{A}} ( \eta' \bar{K}^0 ) & = & \frac{1}{\sqrt{6}}
  \mathcal{P}_{\eta'} \left[ 1 + \zeta' e^{i \delta'} + \xi_s e^{i ( \delta_s +
  \phi_s)} \right], \nonumber\\
  \bar{\mathcal{A}} ( \eta' K^- ) & = & \frac{1}{\sqrt{6}} \mathcal{P}_{\eta'}
  \left[ 1 + \chi' e^{i ( \delta' - \varphi )} + \xi_s e^{i ( \delta_s + \phi_s)} \right],  
\end{eqnarray}
where $\delta_s$ and $\phi_s$ are the strong and weak phases generated
by new physics. It's relative size to $\mathcal{P}_{\eta'}$ is denoted
by $\xi_s$.  For simplicity, we assume that the new physics
contribution respects the isospin symmetry under $u \leftrightarrow
d$. This happens to the modes mainly contributing to the QCD penguins
\cite{Kagan:1997qn}. For any specific models such as the two-Higgs-doublet model
\cite{2hdm,wu:1999fe,Wu:2001vq,Wu:2004kr},
the $Z'$ model \cite{Barger:2004hn} etc. the relation between them is computable.  In
the presence of new physics, the expressions for CP asymmetries to
the leading order are modified as follows
\begin{eqnarray}
  a'_0 & \simeq & 2 \zeta' \sin \delta' \sin \gamma - 2 \xi_s \sin \delta_s \sin
  \phi_s , \nonumber\\
  a'_- & \simeq & 2 \chi' \sin ( \delta' - \varphi ) \sin \gamma - 2 \xi_s \sin
  \delta_s \sin \phi_s .
\end{eqnarray}
and the mixing-induced CP asymmetry is given by
\begin{eqnarray*}
  S' & \simeq & \sin 2 \beta + 2 \zeta' \cos 2 \beta \cos \delta' \sin \gamma - 2
  \xi_s \cos 2 \beta \cos \delta_s \sin \phi_s .
\end{eqnarray*}
Note that in this case $R'$ is not affected as the new physics contributions to
charged and neutral modes cancel. The difference between two direct
CP asymmetries $a'_0 - a'_-$ is not affected either. In the presence
of new physics, the weak phase $\gamma$ extracted from
Eq.(\ref{gamma}) will be an effective one denoted by $\tilde{\gamma}$,
and is related to the true value of $\gamma$ through 
\begin{eqnarray}
  \tan \tilde{\gamma} & \equiv & \frac{1}{r \Delta R'} \left[ ( r - \cos
  \varphi ) \frac{S' - \text{$\sin 2 \beta$}}{\cos 2 \beta} - \sin \varphi
  \cdot a'_0 \right] \nonumber\\
  & \simeq & \tan \gamma - 2 \frac{r \cos \delta_s - \cos ( \delta_s - \varphi
  )}{r \Delta R'} \xi_s \sin \phi_s  .
\label{new-gamma}
\end{eqnarray} 
Thus the deviation of the effective value $\tilde{\gamma}$ from the true
$\gamma$ is a measure of the new physics effects and which can be used to 
extract new physics parameters or distinguish different new physics 
models \cite{Zhou:2000ym}. 
 The true value of $\gamma$ can be obtained from other 
measurements such as through $B\to DK$\cite{Atwood:1996ci} or from global CKM fits.  The
new physics effects will be enhanced if the deviation of $R'$ from
unity is tiny.  As the current data give a central value of $R'\simeq
1.04$, the effective $\tilde{\gamma}$ is very sensitive to new
physics.
If the true value of $\gamma$ is indeed around $62^\circ$,
for typical values of $r$ and $\varphi$ taken from Eq.(\ref{big-r})
and $R'=1.04$, the enhancement factor is about $\sim 50$. 
As a consequence, significant difference of a few tens degree 
between $\tilde{\gamma}$ and $\gamma$  is possible for 
$\xi_s\sin\phi_s \sim 0.1$.

It has been argued recently that in general the new physics will not
generate significant relative strong phases as the strong phases 
mainly originate from the long-distance rescatterings of the final
states while new physics contributes only to short-distant part\cite{Datta:2004re}.
In the case that the new physics strong phase $\delta_s$ is negligible, the
combined new physics parameter $\xi_s \sin \phi_s$ can be directly
extracted.  As an illustration, we take the central value of $r=0.56$
and $\varphi=-33^\circ$ from Eq.(\ref{big-r}) and $R'$ in the
$1\sigma$ range, which gives
\begin{eqnarray}
  0 \alt \xi_s \sin \phi_s \alt 0.25 . 
\end{eqnarray}
It follows that
for a large $r$, the current data marginally agree with the SM,
and  the new physics receives  only an  upper bound. 
 
For a smaller value of $r=0.2$ and $\varphi=-12^\circ$ in
Eq.(\ref{small-r}), a positive signal of nonzero $\xi_s \sin \phi_s$
is found
\begin{eqnarray}
  0.15 \alt \xi_s \sin \phi_s \alt 0.19 ,
\end{eqnarray}
which demonstrates that the $\eta^{(')}K$ mode provide a good avenue
to explore new physics contributions.  Needless to say that the
current experimental status is not conclusive yet and one can not draw
a robust conclusion on the presence of new physics.
% advantage
The advantage of using Eq.(\ref{new-gamma}) in $\eta' K$ modes to
probe new physics is that besides new physics parameters the
difference between the effective and the true $\gamma$ only depends on
the hadronic parameters $r$ and $\varphi$. The knowledge of the
tree-penguin ratio $\zeta^{(')}$ and $\chi^{(')}$ are not needed.
Comparing with  probing new physics through $B_s\to KK$, although the flavor
structure in $B_s\to KK$ is simpler, 
the tree-penguin
interference can not be avoid and one has to combine it with
$B\to\pi\pi$ where additional assumptions on  new physics effects in $b\to d$ penguin
have to be made\cite{London:2004ej}.

\section{conclusion}
In summary, we have present a method for an independent determination of
the weak phase $\gamma$ from $B \rightarrow \eta^{(' )} K$ alone, which
makes use of  measurements of all the direct and mixing-induced
CP asymmetries.
The value of $\gamma$ extracted from $\eta^{(' )} K$ may be compared 
with the ones from other modes. The possible discrepancy may help us
to understand the current puzzles in charmless $B$ decays. 
We have taken two sets of the ratio $\mathcal{C}/\mathcal{T}$ as
inputs to analysis the implications of the recent data on $\eta'K$
 modes. One is from from global $\pi\pi$, $\pi K$ and $K K$ fits
which leads to a $45^\circ \alt \gamma \alt 110^\circ$ in agreement
with the SM fit value. The other is from QCD factorization
calculations which makes $\gamma$  around $90^\circ$.  Within
the SM, it implies that a large $\mathcal{C}$ is independently favored
in $\eta' K$ modes.
New physics beyond SM can be singled out if $\gamma$ obtained in
$\eta' K$ modes is significantly different than the ones from
other decay modes or other approaches. The value of $\gamma$ obtained
from $\eta' K$ are found to be sensitive to new physics contributions
and can be used to extract new physics parameters if the new physics
does not carry significant new strong phases.
%
%One would expect a similar situation should be found in for $\pi K$
%system as there also observed a possible suppresion of $S ( \pi^0 K_S
%)$. Unfurtunately, the tree-penguin interferences in $\pi K$ is much
%more complecated. One can not construct a similar relation like
%Eq.(\ref{gamma}).

%%%%%%%%%%%%%%%%%%%%%%%%%%%-Reference-%%%%%%%%%%%%%%%%%%%%%%%%%%%%%%%%%%%%%%%%%%
%\begin{acknowledgments}
%\end{acknowledgments}
\bibliographystyle{apsrev}
%\bibliography{/home/local/zhou/reflist/reflist,misc}
\bibliography{/home/zhou/reflist/reflist,misc}
\end{document}